\documentclass[twocolumn, 5p, 12pt,fleqn]{elsarticle}
\usepackage{graphicx}
\usepackage{subfigure}
\usepackage[dvipdfx,cmyk]{xcolor}
\usepackage{amssymb}

\biboptions{sort&compress}

\journal{Physics Letters A}

\newcommand{\ds}{\displaystyle}

\newcommand{\ben}{\begin{enumerate}}
\newcommand{\een}{\end{enumerate}}
\newcounter{note}

\newcommand{\vv}[1]{\mathbf{#1}}
\newcommand{\be}{\begin{equation}}
\newcommand{\ee}{\end{equation}}

\begin{document}

\begin{frontmatter}
\title{Symplectic map description
of Halley's comet dynamics}

\author{G.~Rollin}
%\ead{rollin@obs-besancon.fr}

\author{P.~Haag}
%\ead{pierre.haag@edu.univ-fcomte.fr}

\author{J.~Lages}
\ead{jose.lages@utinam.cnrs.fr}
\ead[url]{http://perso.utinam.cnrs.fr/$\sim$lages}

\address{Institut UTINAM, Observatoire des Sciences de l'Univers THETA, CNRS, Universit\'e de Franche-Comt\'e, 25030 Besan\c con, France}

\begin{abstract}
We determine the two-dimensional symplectic map describing 1P/Halley chaotic dynamics.
We compute the Solar system kick function ie the energy transfer to 1P/Halley along one
passage through the Solar system. Each planet contribution to the Solar system kick function
appears to be the sum of a Keplerian potential and of a rotating gravitational dipole potential
due to the Sun movement around Solar system barycenter. The Halley map gives a reliable description
of comet dynamics on time scales of $10^4$yr while on a larger scales the parameters of the map are
slowly changing due to slow oscillations of orbital momentum.
% The main features of 1P/Halley chaotic dynamics can be described by a two dimensional symplectic map.
% Using Melnikov integral we semi-analytically
% determine such a map for 1P/Halley taking into account gravitational interactions from the Sun and the eight planets.
% We determine the Solar system kick function \textit{ie} the energy transfer to 1P/Halley along one passage through the Solar system.
% Our procedure allows to compute
% for each planet its contribution to the Solar system kick function which appears to be the sum of the Keplerian potential of the planet and
% of a rotating circular gravitational dipole potential
% due to the Sun movement around Solar system
% barycenter. We test the robustness of the symplectic Halley map by directly integrating Newton's equations over $\sim2.4\cdot10^4$ yr
% around J2000.0 and by reconstructing
% the Solar system kick function. Our results show that the Halley map
% with fixed parameters gives a reliable description
% of comet dynamics on time scales of $10^4$ yr
% while on a larger scales the parameters of the map
% are slowly changing due to
% slow oscillations of orbital momentum.
\end{abstract}

\end{frontmatter}
%%%%%%%%%%%%%%%%%%%%%%%%%%%%%%%%%%%%%%%%%%%%%%%%%%%%%%%%%%%%%%%%%%%%%%%%%%%%%%%
%******************************************************************************

\section{Introduction}

The short term regularity of 1P/Halley appearances in the Solar system (SS) contrasts with its long term irregular and unpredictable orbital behavior
governed by dynamical chaos \cite{chi89}. Such chaotic trajectories can be described by a Kepler map \cite{petrosky86,chi89}
which is a two dimensional area preserving map involving energy and time. The Kepler map was originally analytically derived in the framework of the
two dimensional restricted three body problem \cite{petrosky86} and numerically constructed for the three dimensional realistic case
of 1P/Halley \cite{chi89}. Then the Kepler map has been used to study nearly
parabolic comets with perihelion beyond Jupiter orbital radius \cite{petrosky86,petrosky88,liu94,shevchenko11}, 1P/Halley chaotic dynamics
\cite{chi89,dvorak90},
mean motion resonances with primaries \cite{malyshkin99,pan04}, chaotic diffusion of comet trajectories
\cite{emelyanenko92,malyshkin99,zhou00,zhou01,zhou02} and chaotic capture of dark matter by the SS and galaxies
\cite{khriplovich09,lages13,rollin14}. Alongside its application in celestial dynamics and astrophysics the Kepler map has been also
used to describe atomic physics phenomena such as microwave ionization of excited hydrogen atoms \cite{casati87,casati90,shepelyansky12},
and chaotic autoionization of molecular Rydberg states \cite{benvenuto94}.

In this work we semi-analytically determine the symplectic map describing 1P/Halley dynamics taking into account the Sun and the eight major planets of the SS.
We use Melnikov integral (see \textit{eg} \cite{liu94,chi79,zaslavsky1,zaslavsky2,reichl,afraimovich}) to compute exactly the kick functions associated to each major planet and in particular we retrieve the kick
functions of Jupiter and Saturn which were already numerically extracted by Fourier analysis \cite{chi89} from previously observed and computed 1P/Halley
perihelion passages \cite{yeomans81}. We show that each planet contribution to the SS kick function can be split into a Keplerian potential term and a
rotating dipole potential term due to the Sun movement around SS barycenter. We illustrate the chaotic dynamics of 1P/Halley with the help
of the symplectic Halley map and give an estimate of the 1P/Halley sojourn time. Then we discuss its long term robustness comparing
the semi-analytically computed SS kick function to the one we extract from 
an exact numerical
integration of Newton's equation for Halley's comet orbiting the SS constituted by the eight planets and the Sun
(see snapshots in Fig.~\ref{fig1}) from -1000 to +1000 Jovian years around J2000.0 \textit{ie} from about -10 000BC to about 14 000AD.
Exact integration over a greater time interval does not provide exact ephemerides since Halley's comet dynamics is chaotic, see eg   
\cite{dvorak90} where integration of the dynamics of SS constituted by the Sun, Jupiter and Saturn have been computed for $10^6$ years.

\section{Symplectic Halley map}

\begin{figure}[t] 
\begin{center} 
\includegraphics[width=\columnwidth]{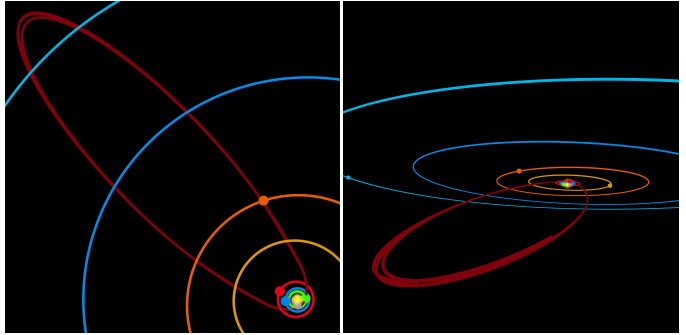}
\caption {\label{fig1}
Two examples of three dimensional view of Halley's comet trajectory. 
The left panel presents an orthographic projection and the right panel presents an arbitrary point of view. The red trajectory shows three successive
passages of Halley's comet through SS,
the other near circular elliptic trajectories are for the eight
Solar system planets, the yellow bright spot gives the Sun position.
At this scale details of the Sun trajectory is not visible.
}
\end{center}
\end{figure}

Orbital elements of the current osculating orbit of 1P/Halley are \cite{jpl}
\begin{center}
\begin{tabular}{lllllll}
$e$&$\simeq$&$0.9671$,&
$q$&$\simeq$&$0.586$ au,\\
$i$&$\simeq$&$162.3$,&
$\Omega$&$\simeq$&$58.42$,\\
$\omega$&$\simeq$&$111.3$,&
$T_0$&$\simeq$&$2446467.4$ JD
\end{tabular}
\end{center}
Along this trajectory (Fig. \ref{fig1}) the comet's energy per unit of mass is $E_0=-1/2a=(e-1)/2q$ where
$a$ is the semi-major axis of the ellipse. In the following we set the gravitational constant
$G=1$, the total mass of the Solar system (SS)
equal to 1, and the semi-major axis of Jupiter's trajectory equal to 1. In such units
we have $q\simeq0.1127$, $a\simeq3.425$ and $E_0\simeq-0.146$. Halley's comet pericenter
can be written as $q=a\left(1-e\right)\simeq\ell^2/2$ where $\ell$ is the intensity per unit
of mass of the comet angular momentum vector. Assuming that the latter changes
sufficiently slowly in time we can consider the pericenter $q$ as constant for many comet's
passages through the SS. We have checked by direct integration of Newton's equations
that this is actually the case ($\Delta q\simeq0.07$) at least for a period of -1000 to +1000 Jovian years around
J2000.0. Consequently, Halley's comet orbit can be reasonably characterized by its semi-major axis $a$
or equivalently by Halley's comet energy $E$. During each passage through the SS
many body interactions with the Sun and the planets modify the comet's energy. The successive
changes in energy characterize Halley's comet dynamics.

Let us rescale the energy $w=-2E$ such as now positive energies ($w>0$) correspond to elliptic orbits and negative energies ($w<0$) to hyperbolic orbits.
Let us characterize the $n$th passage at the pericenter by the phase $x_n=t_n/T_J\, \textrm{mod}\, 1$ where $t_n$ is the date of the passage and $T_J$ is Jupiter's
orbital period considered as constant. Hence, $x$ represents an unique position of Jupiter on its own trajectory.
The energy $w_{n+1}$ of the osculating orbit after the $n$th pericenter passage is given by
\be\label{map}
\begin{array}{ccccc}
w_{n+1}&=&w_n&+&F(x_n)\\ 
x_{n+1}&=&x_n&+&w_{n+1}^{-3/2}
\end{array}
\ee
where $F(x_n)$ is the kick function, \textit{ie} the energy gained by the comet during the $n$th passage and depending on Jupiter phase $x_n$ when the comet is at pericenter.
The second row in (\ref{map}) is the third Kepler's law giving the Jupiter's phase at the $(n+1)$th passage from the one at the $n$th passage and the energy of
the $(n+1)$th osculating orbit.

The set of equations (\ref{map}) is a symplectic map which captures in a simple manner the main features of Halley's comet dynamics. This map has already
been used by Chirikov and Ve\-ches\-la\-vov \cite{chi89} to study Halley's comet dynamics from previously observed or computed perihelion passages from $-1403$BC to
$1986$AD\cite{yeomans81}.
In \cite{chi89} Jupiter's and Saturn's contributions to the kick function $F(x)$ had been extracted using Fourier analysis.
In the next section we propose to semi-analytically compute the exact contributions of each of the 
eight SS planets and the Sun.

\section{Solar system kick function}

Let us assume a SS constituted by eight planets with masses $\left\{\mu_i\right\}_{i=1,\dots,8}$ and the
Sun with mass $1-\mu=1-\sum_{i=1}^8\mu_i$. The total mass of the SS is 
set to $1$ and $\mu\ll1$. In the barycentric reference frame we assume that the eight planets have nearly circular elliptical trajectories with 
semi-major axis $a_i$. We rank the planets such as $a_1<a_2<\dots<a_8$ so $a_5$ and $\mu_5$ are the orbit semi-major axis and the mass of Jupiter.
The corresponding mean planet velocities $\left\{v_i\right\}_{i=1,\dots,8}$ are such as 
$v_i^2=\left(1-\sum_{j\geq i}\mu_j\right)/a_i\simeq1/a_i$. Here we have set the gravitational constant $G=1$
and in the following we will take the mean velocity of Jupiter $v_5=1$. The Sun trajectory in the
barycentric reference frame is such as $\left(1-\mu\right)\vv r_\odot=-\sum_{i=1}^8\mu_i\vv r_i$.

In the barycentric reference frame, the potential experienced by the comet is consequently
\begin{eqnarray}
\Phi(\vv{r})&=&-\ds\frac{1-\mu}{\|\vv r-\vv r_\odot\|}-\ds\sum_{i=1}^8\frac{\mu_i}{\|\vv r-\vv r_i\|}\nonumber\\
&=&\Phi_0(r)\Bigg[1\nonumber\\
&+&\ds\sum_{i=1}^8\mu_i\left(-1-\ds\frac{\vv r\cdot\vv r_i}{r^2}+\ds\frac{r}{\|\vv r-\vv r_i\|}\right)\Bigg]\nonumber\\
&+&o\left(\mu^2\right)\label{phir}
\end{eqnarray}
where $\Phi_0(\vv r)=-1/r$ is the gravitational potential assuming all the mass is located at the barycenter.

Let us define a given osculating orbit $\mathcal{C}_0$ with energy $E_0$
and corresponding to the $\Phi_0(r)$ potential. The change of
energy for the comet following the osculating orbit $\mathcal{C}_0$ under the influence of
the SS potential $\Phi(\vv r)$ (\ref{phir}) is given by the integral
\begin{equation}\label{Melnikov}
\!\!\!\!\!\Delta E\left(x_1,\dots,x_8\right)=\ds\oint_{\mathcal{C}_0}\nabla\left(\Phi_0(r)-\Phi(\vv r)\right)\cdot d\vv r
\end{equation}
which gives at the first order in $\mu$
\begin{eqnarray}
\label{DE}
&&\!\!\!\!\!\!\!\!\!\!\!\!\Delta E\left(x_1,\dots,x_8\right)\nonumber\\
&&\!\!\!\!\!\!\!\!\!\!\!\!\simeq\ds\sum_{i=1}^8\mu_i\oint_{\mathcal{C}_0}\vv \nabla\left(\ds\frac{\vv r\cdot\vv r_i}{r^3}
-\ds\frac{1}{\|\vv r-\vv r_i\|}\right)\cdot d\vv r\\
&&\!\!\!\!\!\!\!\!\!\!\!\!\simeq\ds\sum_{i=1}^8\Delta E_i\left(x_i\right)\nonumber
\end{eqnarray}
This change in energy depends on the phases $\left(x_i=t/T_i\,\textrm{mod}\,1\right)$ of the planets when the comet passes through pericenter.
From (\ref{DE}) we see that each
planet contribution $\Delta E_i\left(x_i\right)$ are decoupled from the others and can be computed separately.

The integral (\ref{Melnikov}) is similar to the Melnikov integral (see eg \cite{liu94,chi79,zaslavsky1,zaslavsky2,reichl,afraimovich}) which
is usually used in the vicinity of the separatrix to obtain the energy change
of the pendulum perturbed by a periodic parametric term.
%Close to the separatrix the perturbed pendulum dynamics can be described by the whisker map \cite{chi79}.
In the case of the restricted 3-body problem the Melnikov integral can be used to obtain the energy change of the light body in the vicinity of
2-body parabolic orbit ($w\simeq0$) \cite{liu94}.
%In that case the light body dynamics can be described by the Kepler map (\ref{map}). The main
%difference between the Kepler map and the whisker map is the way their phase equation diverges for $w\simeq0$, algebraically and
%logarithmically respectively. Chaos is consequently stronger in the vicinity of the parabolic orbit than of the separatrix \cite{petrosky86}.
We checked that integration (\ref{Melnikov}) along an elliptical osculating orbit or along the parabolic orbit corresponding to the same pericenter
give no noticeable difference as long as the comet semi-major axis is greater than planet semi-major axis. To be more realistic we adopt integration
over an elliptical osculating orbit $\mathcal{C}_0$ since in the case of 1P/Halley slight differences start to appear for Neptune contribution to the
kick function.

% \begin{figure}[t] 
% \begin{center} 
% \includegraphics[width=\columnwidth]{fig2.eps}
% \caption {\label{fig2}
% ($a$) Jupiter and ($b$) Saturn contributions, $F_5(x)$ and $F_6(x)$, to the SS kick function
% $F(x)$ (\ref{map}) obtained
% from Melnikov integral calculation (thick line) and extracted by Fourier analysis from
% observations and exact numerical calculus (\textcolor{red}{$\bullet$}, see Fig. 2a in 
% \cite{chi89}). The red dashed line (the blue dotted-dashed line)
% shows the Keplerian contribution (the dipole contribution) to
% the Melnikov integral.
% }
% \end{center}
% \end{figure}

After the comet's passage at the pericenter, when the planet phases are $x_1,\dots,x_8$, the new osculating orbit corresponds to the energy
$E_0+\Delta E\left(x_1,\dots,x_8\right)$. Knowing the relative positions of the planets, the knowledge of \textit{eg} $x=x_5$ is sufficient to determine all the $x_i$'s.
Hence, for Halley map (\ref{map}) the kick function of the SS is $F(x)=-2\Delta E\left(x\right)=\sum_{i=1}^8F_i(x_i)$ where $F_i(x_i)$ is the kick function of the $i$th
planet. In the following we present results obtained from the computation of the Melnikov integral (\ref{Melnikov}) using coplanar circular trajectories for planets.
We have
checked the results are quite the same in the case of the non coplanar nearly circular elliptic trajectories for the planets taken at J2000.0 (see dashed lines in Fig.~\ref{fig3} right panel).

\begin{figure}[!ht] 
\begin{center} 
\includegraphics[width=\columnwidth]{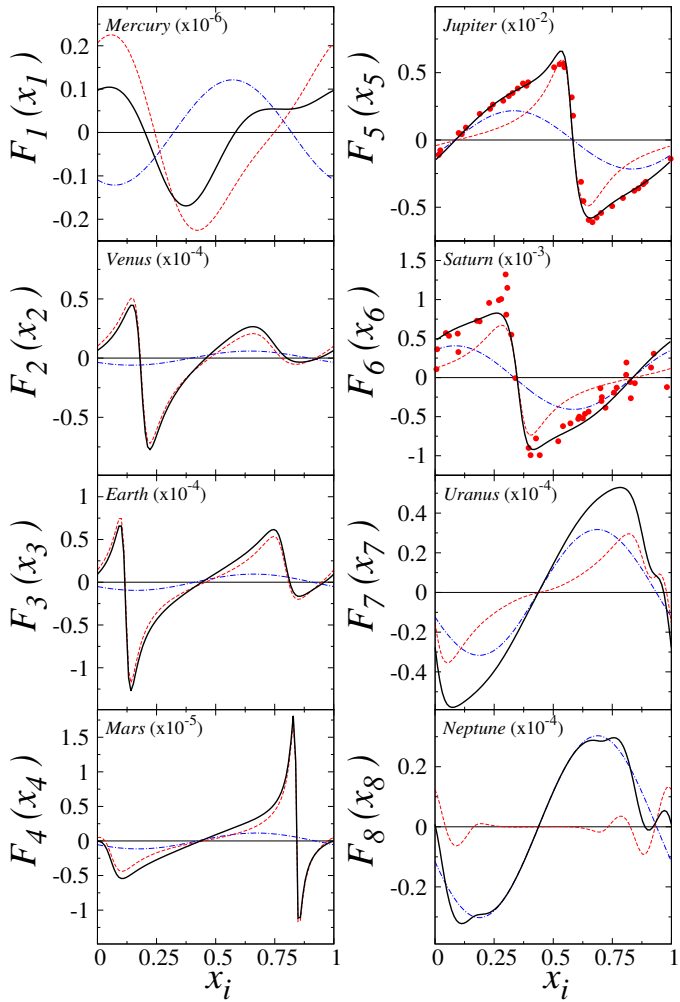}
\caption {\label{fig2}
Contributions of the eight planets to SS kick
function $F(x)$. On each panel $F_i(x_i)$ is obtained from
Melnikov integral calculation (thick line), the red dashed line
(the blue dotted-dashed line)
shows the Keplerian contribution (dipole contribution) to
the Melnikov integral.
On Jupiter and Saturn panels, the kick functions extracted by Fourier analysis from
observations and exact numerical
calculations are shown (\textcolor{red}{$\bullet$}, see Fig. 2
in \cite{chi89}).
}
\end{center}
\end{figure}

Fig.~\ref{fig2} shows contributions for each of the eight planets to the SS kick function.
The two uppermost panels in Fig.~\ref{fig2} right column show contributions of Jupiter, $F_5(x)$, and Saturn, $F_6(x_6)$, to the SS kick function.
We set $x=x_5=0$ when Halley's comet was at perihelion in 1986.
We clearly see that the exact calculus of the Melnikov integral (\ref{Melnikov})
are in agreement with the contributions of Jupiter and Saturn extracted by Fourier analysis \cite{chi89} of previously observed and computed perihelion passages \cite{yeomans81}.
As seen in Fig.~\ref{fig2} the kick function is the sum of two terms (\ref{DE}): the Kepler potential term $-\|\vv r-\vv r_i\|^{-1}$ (dashed red line in Fig. \ref{fig2})
and the dipole potential term $\vv r\cdot\vv r_i/r^3$ (dot dashed blue line in Fig. \ref{fig2}).
These two terms are of the same order of magnitude, the dipole term due to the Sun displacement around the SS
barycenter is therefore not negligible for Jupiter (Saturn) kick function.
The rotation of the Sun around SS barycenter
creates a rotating circular dipole of amplitude $\mu_i\simeq M_i/M_S$
similar to the one analyzed for Rydberg molecular states \cite{benvenuto94}
that gives additional kick function of sinus form.

In Fig.~\ref{fig2} we clearly see that
the saw-tooth shape used in \cite{chi89} to model the kick function is only a peculiar characteristic of Jupiter and Saturn contributions.
Also, the sinus shape analytically found in \cite{petrosky86} for large $q$ can only be considered as a crude model for the planet contributions of the SS kick function.
% According to (\ref{DE}) each contribution $F_i(x)$ is proportional to $\mu_i$
% the ratio between the $i$th planet mass and the total SS mass which explains the different kick function magnitude observed in Fig.~\ref{fig2}.
%As Venus, Earth and Mars
%have semi-major axis of the order of Halley's comet perihelion, the associated kick functions are peaked for phases $x$ giving large transfer of energy.
For Venus, Earth and Mars
%those planets
the kick function is dominated by the Kepler potential term, the dipole potential term being weaker by an order of magnitude. 
Uranus contribution to the SS kick function share the same characteristics as Jupiter's (Saturn's) contributions but two (one) orders of magnitude weaker.
For Neptune as its semi-major axis is
about 60 times greater than Halley's comet perihelion, the direct gravitational interaction of Neptune is negligible and the dipole term dominates the kick function.
Neptune indirectly interacts on Halley's comet by influencing the Sun's trajectory.
As Mercury semi-major axis is less than perihelion's comet, Mercury, like the Sun, acts as a second rotating dipole, consequently the two potential terms in (\ref{DE})
contribute equally.

\begin{figure}[!t] 
\begin{center} 
\includegraphics[width=\columnwidth]{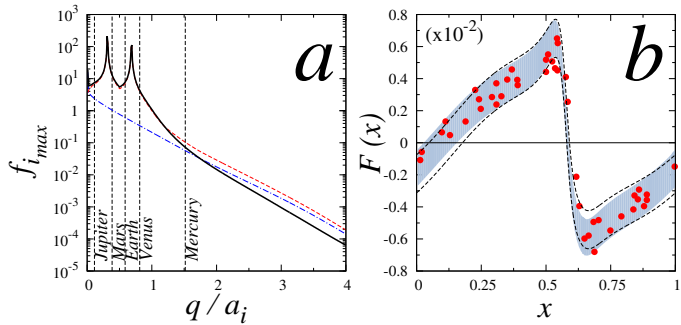}
\caption {\label{fig3}
Left panel: Peak amplitude of the kick function shape $f_i(x_i)$ (thick black line) as a function of pericenter distance $q/a_i$  \cite{sm21}.
The red dashed line (the blue dotted-dashed line) shows the maximum amplitude of the Keplerian contribution (dipole contribution).
Vertical dashed lines show relative positions of planets. On that scale Saturn, Uranus and Neptune relative positions are not shown.
Right panel: Variation domain of the SS kick function $F(x)$ (light blue shaded area)
as a function of Jupiter's phase $x=x_5$. The variation width
is $\Delta F\simeq0.00227$. Data from observations
and exact numerical calculations (Fig. 1 from \cite{chi89}) are
shown (\textcolor{red}{$\bullet$}). The dashed lines bound the variation
domain of the SS kick function when current elliptical trajectories for planets are considered.
}
\end{center}
\end{figure}

For a given osculating orbit, the shape $f_i(x_i)$ of the kick function defined such as $F_i(x_i)=\mu_i f_i(x_i) v_i^2$ has to be only
dependent on $q/a_i$. Let us use the case of 1P/Halley to study general features of $f_i(x_i)$.
Fig.~\ref{fig3} left panel shows the peak amplitude ${f_i}_{max}$ of $f_i(x_i)$.
In the region $0.25\lesssim q/a_i\lesssim0.75$
the peak amplitude ${f_i}_{max}$ is clearly dominated by the Keplerian potential term and even diverges for close encounters at $q\simeq0.3a_i$ and $q\simeq0.7a_i$.
For $q\gtrsim1.5a_i$ the Keplerian potential and the circular dipole potential terms give comparable sine waves almost in phase opposition
(Fig.2 top left panel and \cite{sm21}). We clearly observe for $q\gtrsim1.5a_i$ an exponential decrease of the peak amplitude,
${f_i}_{max}\sim \exp(-2.7q/a_i)$, consistent with the two dimensional case studied in \cite{petrosky86,petrosky88}.

The orbital frequency of the planets being only near integer ratio, for a sufficiently long time randomization occurs and any 8-tuple $\{x_i\}_{i=1,\dots,8}$
can represent the planets position in
the SS. For $x=x_5$ the SS kick function $F(x)$ is a multivalued function for all $0\leq x\leq1$. We can nevertheless define a lower and upper bound to the 
SS kick function which are presented as the boundaries of the blue shaded region in Fig.~\ref{fig3} right panel. We clearly see that raw data points extracted in \cite{chi89}
from previously observed and computed Halley's comet passages at perihelion \cite{yeomans81} lie in the variation domain of $F(x)$ deduced from the Melnikov integral (\ref{Melnikov}).

% \begin{figure}[t] 
% \begin{center} 
% \includegraphics[width=\columnwidth]{fig4.eps}
% \caption {\label{fig4}
% Variation domain of the SS kick function $F(x)$ (light blue shaded area)
% as a function of Jupiter's phase $x=x_5$. The variation width
% is $\Delta F\simeq0.00227$. Data from observations
% and exact numerical calculations (Fig. 1 from \cite{chi89}) are
% shown (\textcolor{red}{$\bullet$}). The dashed lines bound the variation
% domain of the SS kick function when current elliptical trajectories for planets are considered.
% }
% \end{center}
% \end{figure}

\begin{figure}[t] 
\begin{center} 
\includegraphics[width=\columnwidth]{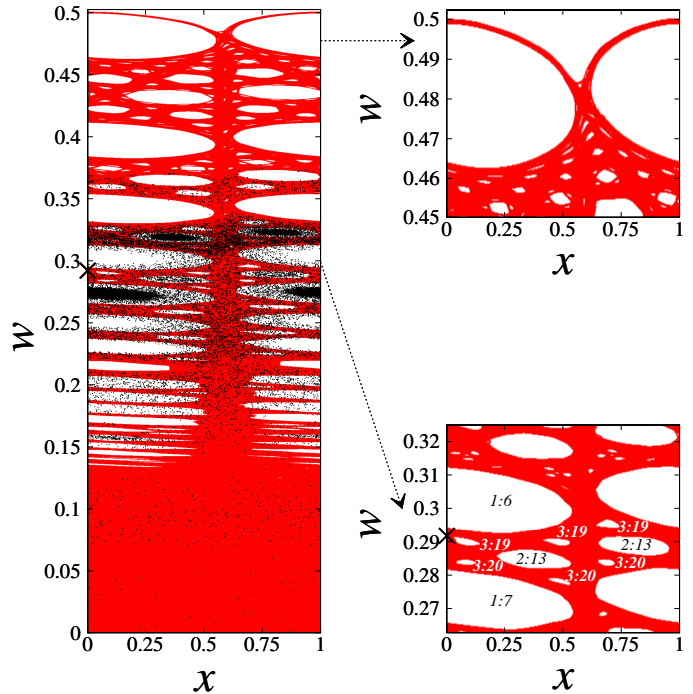}
\caption {\label{fig4}
Left panel: Poincar\'e section of Halley's map generated
only by Jupiter's kick contribution $F_5(x)$ (red area). The cross symbol
($\times$) at $(x=0,w\simeq0.2921)$ gives Halley's comet
state at its last 1986 perihelion passage.
An example of orbit generated by the Halley map (\ref{map}) with the contributions of all the planets is shown by black dots.
Right top panel: closeup on the invariant KAM curve stopping chaotic diffusion.
Right bottom panel: closeup centered on Halley's current
location. Stability islands are tagged with the corresponding
resonance $p$:$n$ between Halley's comet and Jupiter orbital movements.
}
\end{center}
\end{figure}

\section{Chaotic dynamics of Halley's comet}

The main contribution to the SS kick function $F(x)$ is $F_5(x)$ the one from Jupiter as the other planet contributions are from 1 (Saturn) to 4 (Mercury) orders of magnitude weaker.
The dynamics of
Halley's comet is essentially governed by Jupiter's rotation around the SS barycenter. The red shaded area on Fig.~\ref{fig4} shows the section of Poincar\'e obtained from Halley map (\ref{map})
taking only into account Jupiter's contribution $F(x)=F_5(x)$. We clearly see that the accessible part of the phase space is densely filled which
is a feature of dynamical chaos. In the region $0<w\lesssim w_{cr}\simeq0.125$ the comet can rapidly diffuses through a chaotic sea 
whereas in the sticky region $w_{cr}\simeq0.125\lesssim w \lesssim 0.5$
the diffusion is slowed down by islands of stability. The estimated threshold $w_{cr}\simeq0.125$ is the same as the one estimated analytically
in the saw-tooth shape approximation in \cite{chi89}.
Stability islands are located far from the separatrix ($w=0$) on energies
corresponding to resonances
with Jupiter. The current position of Halley's comet $(x=0,w\simeq0.2921)$ is between two stability islands associated with 1:6 and 3:19 resonances
with Jupiter orbital movement (Fig.~\ref{fig4} right bottom panel). As the comet's dynamics is chaotic the unavoidable imprecision on the current
comet energy $w$ allows us only to follow its trajectory
in a statistical sense.
According to the Poincar\'e section (Fig.~\ref{fig4}) associated with Halley map (\ref{map}) for $F=F_5$ the motion of the comet is constrained by a KAM invariant curve
around $w\simeq0.5$ (Fig.~\ref{fig4} top right panel)
constituting an upper bound to the chaotic diffusion.
Consequently as the comet dynamics is bounded upwards the comet will be ejected outside SS as soon as $w$ reaches a negative value.
Taking $10^5$ random initial conditions in an elliptically shaped area with semi-major axis $\Delta x=5\cdot10^{-3}$ and
$\Delta w=5\cdot10^{-5}$ centered at the current Halley's
comet position $(x=0,w=0.2921)$ we find a mean sojourn time of $\overline{\tau}\simeq4\cdot10^8$ yr and a mean number of kicks
of $\overline{N}¡\simeq4\cdot10^4$. A wide dispersion has been observed since $3\cdot10^5$ yr
$\lesssim\tau\lesssim3\cdot10^{13}$ yr and $749\leq N\lesssim9\cdot10^7$.

Now let us turn on also the other planets contributions. As shown in \cite{dvorak90}, where only Jupiter and Saturn are considered,
diffusion inside previously depicted stability islands is now allowed as the other planets act as a perturbation
on the Jupiter's kick contribution. In the example presented in Fig.~\ref{fig4} left panel the comet is locked for a huge number of successive kicks
in a 1:7 and 2:11 resonances with Jupiter around $w\simeq0.27$ and $w\simeq0.32$.
We have also checked that for some other initial conditions even close to the previous example one the KAM invariant curve around $w\simeq0.5$ associated with
the Jupiter contribution (see Fig.~\ref{fig4} top right panel)
no more stops the diffusion towards $w\sim 1$ region where the kicked picture and therefore the map description are no more valid. Taking statistically
the same conditions as in the only Jupiter contribution case we discard about $11\%$ of the initial conditions giving orbits exploring the region $w>0.5$
and for the remaining initial conditions we obtain a mean sojourn time of $\overline{\tau}'\simeq4\cdot10^7$ yr and a mean
number of kicks of $\overline{N}'\simeq3\cdot10^4$. A wide dispersion has been observed since $1\cdot10^5$ yr
$\lesssim\tau'\lesssim6\cdot10^{11}$ yr and $559\leq N'\lesssim5\cdot10^5$.
The two maps give comparable mean number of kicks $\overline{N}'\sim \overline{N}$ but the mean sojourn time is ten time less in the case of
the all-planets Halley map ($\overline{\tau}\sim10\;\overline{\tau}'$). This is due to the fact that the comet can be locked in for a great number
of kicks in Jupiter resonances at large $0.5\gtrsim w\gtrsim0.125$ which correspond to small orbital periods.
In accordance with the results presented in \cite{chi89} we retrieve for the mean sojourn time a $10$ factor between the only Jupiter contribution
case and the all planets contribution case (Jupiter and Saturn only in \cite{chi89}). But we note that the mean sojourn times computed here are $10$
times greater than those computed in \cite{chi89} where only $40$ initial conditions have been used.

\begin{figure}[!t] 
\begin{center} 
\includegraphics[width=\columnwidth]{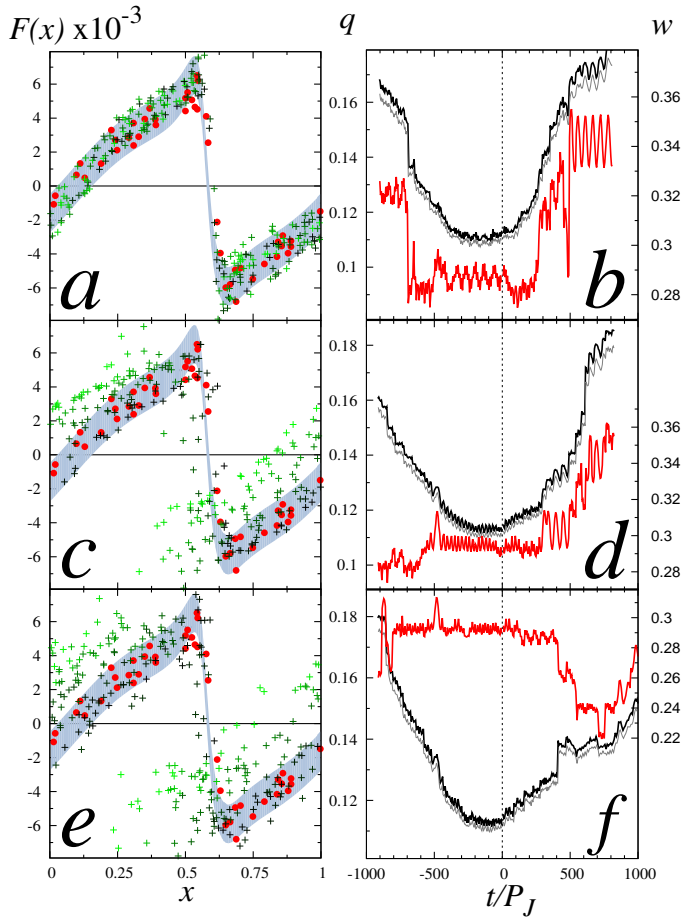}
\caption {\label{fig5}
Numerical simulation of Halley's comet dynamics over
a time period of $-1000$ to $1000$ Jovian years around J2000.0 ($t=0$)
with SS modeled as ($a,b$) the Sun and 8 planets with coplanar circular orbits,
($c,d$) the Sun and Jupiter with non coplanar elliptical orbits, ($e,f$) the Sun
and the eight planets with non coplanar elliptical orbits.
Left panels: kick function $F(x)$ values ($+$) extracted
from $290$ successive simulated pericenter passages of Halley's comet.
The color symbol goes linearly from black for data extracted at time $t=0$
to light green for data extracted at
time $|t|\simeq 10^3P_J$.
We show only points in the range $-0.008<F(x)<0.008$. Data from
observations and exact numerical calculations (Fig. 1 from \cite{chi89})
are shown (\textcolor{red}{$\bullet$}).
Right panels: time evolution of pericenter $q$ (black curves, left
axis)
and of the osculating orbit energy $w$ (red curves, right axis).
Numerical simulations have been done time forward and time backwards
from $t=0$. The gray curves show the time evolution of the $\ell^2/2$
quantity.
}
\end{center}
\end{figure}

\section{Robustness of the symplectic map description}

In order to test the robustness of the kicked picture for Halley's comet dynamics we have directly integrated Newton's equations
for a period of -1000 to +1000 Jovian years around J2000.0 in the case of a SS constituted by the Sun and the eight planets with coplanar circular orbits (Fig.~\ref{fig5} first row),
the Sun and Jupiter with elliptical orbits (Fig.~\ref{fig5} second row), and the Sun and the eight planets with elliptical orbits (Fig.~\ref{fig5} third row).
From Fig.~\ref{fig5} right panels we see that our modern era is embedded in a time interval $-400P_J<t<200P_J$ ($-2800$BC$<t<4400$AD)
with quite constant Halley's comet energy $w\simeq0.29$ and perihelion $q\simeq0.11$. This relatively dynamically quiet time interval allows the good agreement between our
semi-analytic determination of SS kick function using the Melnikov integral (\ref{Melnikov}) and the SS kick function extracted \cite{chi89} from previously observed
and computed perihelion passages \cite{yeomans81}.

In Fig.~\ref{fig5} left panels we reconstruct as in \cite{chi89} the kick function using the dates $t_n$ of the Halley's comet passages at perihelion
$F(x_n)=\left(t_{n+1}-t_n\right)^{-2/3}-\left(t_{n}-t_{n-1}\right)^{-2/3}$. We clearly see that these kick function values lie in the variation domain of the SS kick function when
coplanar circular orbits are considered
for the Sun and the planets (Fig.~\ref{fig5}$a$). In the case of non coplanar elliptical orbits (Fig.~\ref{fig5}$c$ and $e$) the agreement is good but weaker than
the coplanar circular case. This is due to Halley's comet precession which introduces a phase shift in $x$ (see gradient from black to green color in Fig.~\ref{fig5}$a$, $c$ and $e$).
As the coplanar circular orbits case possesses an obvious rotational symmetry, it is
much less affected by the comet precession (Fig.~\ref{fig5}$a$).

In Fig.~\ref{fig5} left panels we show only kick function values in the interval range $-0.008<w<0.008$ corresponding to the variation range of the SS kick function (Fig.~\ref{fig3} right panel)
obtained using Melnikov integral (\ref{Melnikov}). Around the sharp variation $x\simeq0.6$ we obtained few kick function values outside this energy interval
(up to $|F|\simeq0.05$) which corresponds to big jumps
in energy (eg at $t\simeq400P_J$ in Fig.~\ref{fig5}$f$) shown in Fig.~\ref{fig5} right panels. We have checked that those big jumps
occur when Halley's comet at its perihelion approaches closer to Jupiter. As a consequence the two dimensional Halley map (\ref{map}) can be used with confidence only
for short intervals of time $\Delta t\lesssim10^4$yr such as eg the one at $-400P_J\lesssim t\lesssim200P_J$ in Fig.~\ref{fig5} right panels.

\section{Conclusion}

We have exactly computed the energy transfer from the SS to 1P/Halley and we have derived the corresponding symplectic map which characterizes 1P/Halley
chaotic dynamics. With the use of Melnikov integral, energy transfer contributions from each SS planets have been isolated. In particular,
we have retrieved the kick functions
of Jupiter and Saturn previously extracted by Fourier analysis \cite{chi89}. The Sun movement around SS barycenter induces a rotating gravitational
dipole potential which is non negligible in the energy transfer from the SS to 1P/Halley. The symplectic Halley map allows us to follow the chaotic
trajectory of the comet during relatively quiet dynamical periods $\Delta t\lesssim10^4$yr exempt of closer approach with major planets. 
One can expect that a higher dimensional symplectic
map involving the angular momentum and other orbital elements would allow to follow Halley's comet dynamics for longer periods taking into account
large variation in energy (close planet approach) and precession.
In spite of the slow time variation of the Halley map parameters such a symplectic map description allows to get a physical understanding of the global
properties of comet dynamics giving a local structure of phase pace and a diffusive time scale of chaotic escape of the comet from the Solar
system.

\section*{Acknowledgments}

The authors thank D. L. Shepelyansky for useful comments on the current work.

\end{document}